\documentclass[11pt]{article}
\usepackage{amsmath, amssymb}
\begin{document}
\newcommand{\FR}{\mathfrak{R}} \newcommand{\Fh}{\mathfrak{h}}
\newcommand{\FU}{\mathfrak{U}} \newcommand{\FI}{\mathfrak{I}}
\newcommand{\FA}{\mathfrak{A}} \newcommand{\Fa}{\mathfrak{a}}
\newcommand{\FC}{\mathfrak{C}} \newcommand{\FF}{\mathfrak{F}}
\newcommand{\UC}{{\cal U}({\FC})} \newcommand{\UF}{{\cal U}(\widehat{\FF})}
\newcommand{\SR}{{\cal S}(\FR)} \newcommand{\SC}{{\cal S}(\RD \oplus \FR)}
\newcommand{\UR}{{\cal U}(\HR)} \newcommand{\SF}{{\cal S}(\FF)}
\newcommand{\ZZ}{{\cal Z}} \newcommand{\RT}{{\cal R}^t}
\newcommand{\BB}{{\cal B}} \newcommand{\PP}{{\cal P}}
\newcommand{\HU}{{\widehat\FU}} \newcommand{\HR}{{\widehat\FR}}
\newcommand{\Ha}{{\widehat\Fa}} \newcommand{\Hh}{{\widehat\Fh}}
\newcommand{\BR}{{\mathbb{R}}} \newcommand{\RD}{\BR^D}
\newcommand{\BP}{{\mathbb{P}}} \newcommand{\BQ}{{\mathbb{Q}}}
\newcommand{\MMU}{{\mu_1\dots\mu_N}}
\newcommand{\Tr}{\hbox{Tr }}  \newcommand{\Ker}{\hbox{Ker }}
\newcommand{\Eins}{{\mathbf 1}}
\newcommand{\QED}{\hspace*{\fill}Q.E.D.\vskip2mm}
\renewcommand{\theequation}{\thesection.\arabic{equation}}
\newcommand{\shuf}[2] {\boxed{\genfrac{}{}{0pt}{1}{#1}{#2}}}
\renewcommand{\today}{}
\title{\vskip-11mm \bf Algebraic Quantization \\ of the Closed Bosonic String
\footnote{The article is based on the diploma thesis of the first author
(C.M.) \cite{M} under the supervision of K. Pohlmeyer, Universit\"at
Freiburg, completing a project by the second author (K.-H.R.) lying
dormant since around 1987.}}  
\author{{\sc Catherine Meusburger} \\ Fakult\"at f\"ur Physik,
Universit\"at Freiburg, \\ 79104 Freiburg, Germany\thanks{Address
  after January 2002: Dept. of Mathematics, Heriot-Watt University,
  Riccarton, Edinburgh EH14 4AS, UK, electronic address: 
  \tt C.Meusburger@ma.hw.ac.uk}
\\[1mm] and \\[1mm] {\sc Karl-Henning Rehren} \\
Institut f\"ur Theoretische Physik, Universit\"at G\"ottingen,
\\ 37073 G\"ottingen, Germany\thanks{Electronic address:
{\tt rehren@theorie.physik.uni-goe.de}}}

\maketitle

\begin{center}
\vskip-5mm\sl Dedicated to Rudolf Haag on the occasion of his 80th birthday
\end{center}  

\begin{abstract} The gauge invariant observables of the closed bosonic
  string are quantized in four space-time dimensions by
  constructing their quantum algebra in a manifestly covariant
  approach, respecting all symmetries of the classical observables. 
  The quantum algebra is the kernel of a derivation on the universal 
  envelopping algebra of an infinite-dimensional Lie algebra. The search 
  for Hilbert space representations of this algebra is separated from
  its construction, and postponed.\\[5mm]
PACS 2001: 11.25.-w, 11.30.-j, 03.65.Fd \\
MSC 2000: 81R10, 81T30
\end{abstract}

\section{Introduction}
\setcounter{equation}{0} Fock space quantization of String Theory is
notoriously beset with tachyons and anomalies. This fact is
responsible for the need of supersymmetry and extra dimensions whose
introduction and interpretation have triggered several ``string
revolutions'' with a vast range of speculative implications
\cite{GSW}. Yet, these features tend to obliterate the underlying
simple idea of String Theory \cite{NG}. In this article, we prefer to
cling to that simple idea, viewed as a model for a consistent quantum
theory of extended objects with a presumed relevance to gauge theories
\cite{MM,P}, without ambitions towards a Theory of Everything.

The present article demonstrates the viability (with a proviso, 
see below) of an alternative approach to (bosonic closed) string
quantization initiated in \cite{P1} (and pursued in
\cite{PR1,PR2,PR3,P2,P3,P4,HN}) which does not suffer from the
drawbacks mentioned above. This approach differs in at least three
essential aspects from the conventional one. The first is the
interpretation and quantum theoretical implementation of the
constraints \cite{P2}. Second, it strives to capture the observable 
features of the classical theory (and only these) in terms of classically
reparametrization invariant quantities (the observables of the
string), and to quantize only those. A complete system of classical
observables has been identified \cite{PR1,PR3}, see Sect.\ 2 below.

The third substantial distinction of the present approach concerns the
concept of quantization. It is understood in a purely algebraical
sense by consistently promoting Poisson brackets to commutators, that
is, without constructing a Hilbert space representation at the same time. 
The representation problem is thus detached from the construction of
the quantum algebra, and 
this opens the possibility of finding inequivalent physical (positive
energy) representations  of the same quantum algebra (superselection
sectors). The constraints on the classical dynamical variables
\cite{P1,PR2} will be implemented in terms of the appropriate
eigenspaces of Casimir operators of the quantum algebra (rather than
using some elimination prescription). In this way, they determine the
physical representations \cite[and private communication by K. Pohlmeyer]{P2}. 
 
The Poisson algebra $\Fh$ of the classical string observables exhibits
a rich algebraic structure, involving an infinite number of polynomial
relations among multiple Poisson brackets of its generators \cite{PR1,PR2,P3}. 
The presence of these polynomial relations severely complicates the
algebraic construction of the quantum algebra. 

An intrinsic approach to algebraic quantization has been pursued in
\cite{P4,HN}. It assigns a quantum counterpart to each classical
generator, and a quantum polynomial relation to each classical
polynomial relation. One admits observable subleading terms of order
$\hbar$ or higher (quantum corrections) to the relations, restricted by the
grading of the classical algebra. In order to determine the quantum
corrections, one requires a maximum of structural similarity with the
classical algebra; in particular, the commutator of a quantum relation
with an observables must not generate any new relations without a
classical counterpart (the principle of correspondence) \cite{P4}.  

With these postulates and Poincar\'e covariance as guiding principles,
one proceeds degree by degree in the inherent algebraic grading. It
has been shown \cite{P4,HN} that in $1+3$ dimensions and up to degree 5, 
all quantum corrections can be consistently and almost
uniquely determined. Three free parameters survive a highly
overdetermined non-linear system of conditions at this degree.%
\footnote{The complexity of the problem is illustrated by the number
  of 106.089 relations at degree $5$ among 2.337 invariants of degree
  $\leq 5$ (in $1+3$ dimensions) \cite{HN}.}

It remains unsatisfactory, of course, (and impracticable on the long
run), to proceed degree by degree. The aim of the present article is to
establish the existence of a consistent quantization prescription
to all degrees.

We succeed to do so with the only proviso that, in the final step
of our construction, we extrapolate an algebraic feature of the
classical algebra apparent at lower degrees ($\leq 7$), and assume
its persistence to all degrees. Sect.\ 6 is devoted to the discussion
of this ``quadratic generation property'', which would follow from a
structural property of the underlying (explicitly known)
infinite-dimensional Lie algebra. Numerical tests of this linear
problem have found their limitation due the rapid growth of this
graded Lie algebra \cite{P4}, while structural arguments available so
far are only partial \cite{P3,P5}.  

We pursue an extrinsic approach, which is slightly against the spirit
outlined above. The idea of such an approach is to take advantage of
an embedding of the classical observables into an auxiliary ``ambient'' 
Poisson algebra in which (i) the polynomial {\em relations} among the
observables are {\em identities} in terms of independent (but
non-observable) variables, and for which (ii) a standard quantization
procedure is available. As an analogy from mechanics, instructive for
this and the following (though infinitely simpler), the reader is
invited to think of the relation    
$$\{A,B\}^2-16A \cdot B=0$$ 
which turns into an identity if one embeds $A=P^2$, $B=Q^2$ into the
canonical Poisson algebra generated by $P$ and $Q$. Identities in the
classical ambient algebra will acquire quantum corrections of order
$\hbar$ or higher in the quantized algebra. 

The quantum observables are then sought as elements of the 
non-commut\-ative quantized
ambient algebra, e.g., by the specification of an embedding
prescription (such as a suitable factor ordering) for the expression
of a generating set of observables in terms of the non-observable
quantized variables. Thus, classical polynomial relations involving
multiple Poisson brackets of the generators naturally give rise
to the corresponding polynomial relations in the quantum algebra
involving multiple commutators of the generators, exhibiting
quantum corrections. These quantum corrections, of course, a priori
belong to the {\it ambient algebra}.

The principle of correspondence in this case stipulates that they must
be quantum {\it observables}, since otherwise the quantized theory
would have new observables without a classical counterpart. The problem 
is thus to establish the existence of an embedding prescription
such that the quantum corrections to the polynomial relations are
observables. Once this has been achieved, the quantum algebra of
observables is well defined and obeys the principle of correspondence. The
remaining challenge of finding and classifying positive energy
Hilbert space representations concerns only this algebra and not the 
quantized ambient algebra which need not be represented.

In the standard approach, the ambient algebra is obtained by canonical
quantization of the string's Fourier modes in a given parametrization. 
It has been shown, however, that this choice, together with frequency
normal ordering, produces non-invariant quantum corrections to
reparametrization invariant relations \cite{B} (apart from the well
known violation of Poincar\'e covariance in that approach) and thus
violates the principle of correspondence.  

The ambient algebra which we choose in this article, is the
envelopping algebra of an infinite-dimensional Lie algebra (with
respect to the Poisson bracket). Its generators are the components of an
infinite set of Lorentz tensors (``monodromy variables'') which depend
on the string parametrization only through the choice of a reference
point on the string's world surface. 

The classical observables are represented as polynomials in the
independent monodromy variables. It follows that multiple Poisson
brackets among observables, and polynomials therein, are also
polynomials in these variables. The polynomial relations defining the
classical algebra of observables are, when expressed as polynomials in
the monodromy variables, identities. (In fact, this is how the
polynomial relations were originally derived in \cite{PR1,PR2,P3}.)   

The quantum ambient algebra is defined by promoting the classical
Poisson bracket of the Lie algebra to a commutator, and passing to the 
universal envelopping algebra. The quantum algebra of observables is
defined as the kernel of a suitable derivation, acting on this
non-commutative algebra. The quantum counterparts of the classical
observable generators are identified as non-commutative polynomials
annihilated by the derivation. This implies that all their multiple
commutators and non-commutative polynomials therein also belong to the
kernel. In particular, the quantum corrections obtained by replacing
in the classical polynomial relations classical generators by quantum 
generators, Poisson brackets by commutators, and commutative products
by non-commutative ones with any choice of factor ordering, belong to
the kernel and hence indeed are also observables. 

This approach results in a consistent covariant quantization of the
string observables in any dimension, complying with the principle of
correspondence, provided we may take for granted the quadratic
generation property.

\section{Outline of the classical structure}
\setcounter{equation}{0}
With the left-moving and right-moving modes ($\pm$) of every classical
solution $x_\mu(\tau,\sigma)$ of the closed bosonic string in $D$
space-time dimensions, one can associate two Lax pairs \cite{P1}. These
are systems of linear partial differential matrix equations whose
integrability conditions are equivalent to the Nambu-Goto equation of
motion. They involve $D$ arbitrary $n \times n$ matrices $T^\mu$,
$\mu=0\dots D-1$, of arbitrary size $n$. The spectra of the corresponding
monodromy matrices $\phi^\pm_T$ are reparametrization invariant non-local 
functionals of the world surface. The observables of the classical bosonic 
string are obtained by variation with respect to the parameter matrices.

We describe the salient algebraic structures of the classical observables
(invariants). More detailed formula will be given in Sect.\ 4.

\vskip2mm{\bf 2.1. Invariants \cite{P1,PR1,PR3}:}

Since the size $n$ of the parameter matrices $T^\mu$ is arbitrary, there
are infinitely many free parameters. Varying $\Tr (\phi^\pm_T-\Eins_n)=
\sum_{N=1}^\infty \ZZ^\pm_\MMU \Tr T^{\mu_1}$ $\cdots T^{\mu_N}$
with respect to the matrices $T^\mu$, one obtains two infinite systems of
reparametrization invariant observables $\ZZ^\pm_{\mu_1\ldots\mu_N}$
(henceforth called {\bf invariants}). They are explicitly given as
iterated integrals over left- and right-moving modes of the canonical string
variables, $u^\pm_\mu(\tau,\sigma) = p^\mu\pm\partial_\sigma x^\mu$,
\begin{equation}
\ZZ^\pm_{\mu_1\ldots\mu_N} = \mathfrak{z}_N\circ
\int_{\sigma<\sigma_N<\cdots<\sigma_1<\sigma+2\pi}\hbox{\kern-22mm}
d\sigma_1\cdots d\sigma_N \quad
u^\pm_{\mu_1}(\tau,\sigma_1)\cdot\ldots\cdot u^\pm_{\mu_N}(\tau,\sigma_N)
\end{equation}
where $\mathfrak{z}_N$ denotes the sum over the cyclic permutations of
the Lorentz indices. They do not depend on the choice of the reference
point $(\tau,\sigma)$ in the formula.

The invariants are reparametrization invariant functionals of the
string's world surface. Their values, together with the infinitesimal
generators of Lorentz boosts, locally determine the world surface up
to translations in the direction of the string's momentum.

With respect to their canonical Poisson bracket (see below) and their
multiplication as functionals they form a Poisson algebra, the
classical algebra of observables. The invariants $\ZZ^+_\MMU$ and
$\ZZ^-_\MMU$ associated with left and right moving modes, respectively,
constitute two Poisson commuting subalgebras $\Fh^+$ and $\Fh^-$ which
are isomorphic up to a global sign in the structure constants.

The invariants are Lorentz tensors $\ZZ^\pm_\MMU$ of any rank $N\geq 1$.
The vector invariants $\PP_\mu=\ZZ^+_\mu=\ZZ^-_\mu$ (arising at first
order in $T^\mu$) are the components of the total momentum of the
string. These are the only invariants which are common to both systems
$\Fh^\pm$ of invariants. The remaining invariants of either system are
algebraically independent of those of the other system, and each
system involves infinitely many algebraically independent ones. Thus,
the quantum algebra $\Hh^{(-)}$ will be the opposite algebra of
$\Hh^{(+)}$. In the sequel, we describe the left-moving one, and omit
the distinguishing superscript $+$. 

\vskip2mm{\bf 2.2. Monodromy variables (truncated tensors) \cite{P1,PR1,PR3}:}

In order to analyse the structure of the classical algebra of
observables, the invariants were expressed as polynomials in another
set of Lorentz tensors, called truncated tensors in \cite{PR1}. In 
the following, we shall prefer the term {\bf monodromy variables}.
They are generated by the logarithm $\log \Phi_T(\tau,\sigma) =
\sum_{N=1}^\infty \RT_\MMU(\tau,\sigma) T^{\mu_1}\cdots T^{\mu_N}$ of the
monodromy matrix, which (unlike the trace) depends on the
choice of a reference point $(\tau,\sigma)$ on the string's world
surface. Therefore, the monodromy variables $\RT_\MMU$ are not
completely reparametrization invariant. Their dependence on the string
coordinates is given by
\begin{align}
\label{ableit}
\partial_\sigma \RT_\MMU(\tau,\sigma) & =
u_{\mu_1}(\tau,\sigma)\RT_{\mu_2\ldots\mu_N}(\tau,\sigma) - 
u_{\mu_N}(\tau,\sigma)\RT_{\mu_1\ldots\mu_{N-1}}(\tau,\sigma)\nonumber \\
\partial_\tau \RT_\MMU(\tau,\sigma) & =
\gamma(\tau,\sigma)\partial_\sigma\RT_\MMU(\tau,\sigma)\,,
\end{align}
where $u_\mu(\tau,\sigma)$ are the canonical string variables and
$\gamma$ is an arbitrary function, which reflects the choice of a gauge.

The monodromy variables $\RT_\MMU$ at any fixed point $(\tau,\sigma)$
are algebraically independent functionals up to linear relations
involving sums over ``shuffle permutations'' of their indices
according to eq.\ (\ref{linear}) below.

The bracket among the monodromy variables induced by the canonical bracket 
between left and right moving modes violates the Jacobi identity due to
the singular behaviour of the canonical bracket. But it can be modified
so as to restore the Jacobi identity, and without affecting the Poisson
bracket among the invariants (which are polynomials in the monodromy
variables) \cite[p.\ 615]{PR1}. In compact notation for the generating
functionals, the (modified) Poisson bracket is given by \cite{KR}
\begin{align}
\label{modpoissgen}
\{\log \Phi_S&\substack{\otimes\\,}\log \Phi_T\}=\\ &=
\Tr\big([S\cdot T,S-T]\cdot\partial_X\big)\log \Phi_X\vert_{X=S+T} -
[S\cdot T,\log \Phi_S - \log \Phi_T] \nonumber
\end{align}
where the matrices $S^\mu$ and $T^\mu$ are understood as
$S^\mu\otimes \Eins$ and $\Eins \otimes T^\mu$, and the differential
operator $\Tr(Y\cdot\partial_X)$ acts as the substitution of $Y^\mu$
for $X^\mu$ in a derivational manner. For a more explicit formula in
terms of the components, see eq.\ (\ref{modpoiss}) below. Equipped
with this bracket, the linear span $\FR$ of the monodromy
variables $\RT_\MMU$ at any fixed point $(\tau,\sigma)$ is an
infinite-dimensional Lie algebra. Its symmetric envelopping algebra
$\SR$ is therefore canonically a Poisson algebra.

The classical invariants are the reparametrization invariant
polynomials in the monodromy variables $\RT_\MMU(\tau,\sigma)$,
characterized by the condition
\begin{equation}
\label{repinv}
\partial_\sigma \ZZ = 0.
\end{equation}
Each of them is a linear combination of tensor components $\ZZ_\MMU$
of the form given in eq.\ (\ref{zyklpol}). Their linear span $\Fh$ is
in fact closed under multiplication and under Poisson brackets
(extended to $\SR$ by the Leibniz rule). In other words, $\Fh$ is the
Poisson subalgebra of $\SR$ given by 
\begin{equation}
\label{classker}
\Fh = \Ker \partial_\sigma \subset \SR.
\end{equation}

\vskip2mm{\bf 2.3. Homogeneity and grading \cite{PR1}:}

The Poisson algebra $\SR$ possesses two gradings: one with respect to
the order $N\geq 1$ in the spectral parameters $T^\mu$ (i.e., the
total Lorentz {\bf tensor rank} disregarding possible contractions),
and one with respect to the {\bf order} $K\geq 1$ as a polynomial in
the monodromy variables.

Both the rank and the order are additive with respect to the product,
whereas the Poisson bracket reduces the total rank by two and the
total order by 1. The two gradings can be unified to a single grading
with respect to the {\bf degree} $l=N-K-1\geq -1$, which is additive
under the Poisson bracket whereas the product of two elements of
degrees $l$ and $l'$ is of degree $l+l'+1$: 
\begin{equation}
\SR = \bigoplus_{l\geq-1} \SR^{(l)}, \qquad
\begin{array}{l}
\SR^{(l)}\cdot\SR^{(l')} \subset \SR^{(l+l'+1)} \\[2mm]
\{\SR^{(l)},\SR^{(l')}\} \subset \SR^{(l+l')}.
\end{array}\end{equation}
The spaces $\SR^{(l)}$ are spanned by spanned by
all monomials of order $K$ in components of $\RT_\MMU$
with total tensor rank $N$ and $N-K-1 = l$. 

The homogeneous parts and the parts of fixed tensor rank of each
invariant polynomial in the monodromy variables are separately invariant.
The gradings of $\SR$ therefore give rise to gradings of $\Fh$. In
particular, with $\Fh^{(l)}=\Fh \cap\SR^{(l)}$, one has the grading
with respect to the degree
\begin{equation}
\label{grading}
\Fh = \bigoplus_{l\geq-1} \Fh^{(l)},\qquad \begin{array}{l}
\Fh^{(l)}\cdot\Fh^{(l')}\subset\Fh^{(l+l'+1)} \\[2mm]
\{\Fh^{(l)},\Fh^{(l')}\}\subset\Fh^{(l+l')} \end{array}
\end{equation}
The subalgebra $\SR^{(-1)}=\Fh^{(-1)}$ consists of the polynomials in the 
components of the total momentum $\PP_\mu=\RT_\mu=\ZZ_\mu$. Except for 
this algebra of central elements acting on each of them as multipliers, 
the subspaces $\SR^{(l)}$ and $\Fh^{(l)}$ are finite-dimensional.

The components $\PP_\mu$ are central in the Lie algebra $\FR$ and
hence in the Poisson algebra $\Fh$. As a function on the classical
phase space, $\PP_\mu$ takes values in the closed forward light-cone,
with light-like values only on rather singular configurations. In the
sequel, we shall regard the mass square $\PP_\mu\PP^\mu$ as a number
$m^2$, i.e., each algebra is tacitly understood as a quotient by the
relation $\PP_\mu\PP^\mu=m^2$, and $m^2$ is assumed $>0$.%
\footnote{The reason for excluding the massless case (for the time
  being) is that in order to generate the Poisson algebra (Sect.\ 2.4 below),
  it is sometimes necessary to divide by $m^2$. (One may as well add
  an element $m^{-2}$ to the algebra $\SR$ which satisfies the obvious
  relations.) All formulae derived in the center of mass system in the
  previous literature can then be easily cast into covariant form. For
  a manifestly covariant treatment, see \cite{H}.}   

The invariants of degree $l=0$ act by Poisson brackets on $\Fh$ as
infinitesimal Lorentz transformations belonging to the stabilizer
subgroup SO(3) of the vector $\PP_\mu$. Thus the invariants of fixed
degree can be classified according to the representations (multiplets)
of SO(3) and parity, which considerably facilitates the investigation
of the algebra \cite{P4}. 

\vskip2mm{\bf 2.4. Generating invariants and polynomial relations \cite{PR1}:}

An algebraic basis of $\Fh$ containing infinitely many algebraically
independent invariants has been identified \cite[Sect.\ IV]{PR1} with
the aid of the monodromy variables. However, it turns out that there
is no  algebraic basis of invariants whose linear span is a Lie algebra. 
Instead, the Poisson bracket between two basis invariants generically
is a polynomial in basis invariants. Hence the Poisson algebra $\Fh$
of classical string observables is not the symmetric envelopping
algebra of a Lie algebra. 

The grading of $\Fh$ with respect to the degree $l$ suggests investigating 
its generation via the Poisson bracket, such that every invariant can
be represented as a polynomial in the generating invariants and their
multiple Poisson brackets. It is known that, as a Poisson algebra,
$\Fh$ is not generated by its elements of lowest degrees $l\leq 1$
\cite[p.\ 622]{PR1}. But there is considerable evidence (for $m^2>0$,
and at least in $1+2$ and $1+3$ dimensions \cite{P3,P4,P5}) that it can be 
generated by elements which are quadratic in the monodromy variables.  
We shall take this {\bf quadratic generation property (hypothesis)}
for granted in the final step of our construction (Prop.\ 4), and
refer to Sect.\ 6 for a discussion of this issue.

More specifically, in $1+3$ dimensions, the generating invariants
consist of the multiplet $J_1$ of generators of the little group
$SO(3)$ (degree 0), three multiplets $S_1$, $S_2$, $T_2$ of degree 1
and an infinite series $\BB_{(l)}$ of Poisson commuting Lorentz
scalars, one at each odd degree $l$. All these generators are
quadratic polynomials in the monodromy variables.

The generators do not generate $\Fh$ freely. There is an infinite number 
of polynomial relations among multiple Poisson brackets of generating
invariants. These relations are identities in $\SR$ (i.e., as polynomials 
in the monodromy variables, and also as functionals of the canonical
string variables). In particular, they are homogeneous in the degree.  

The polynomial relations constitute a Poisson ideal $\FI$ in the symmetric
envelopping algebra $\SF$ of the free Lie algebra $\FF$ of the generators. 
The classical algebra of string observables $\Fh$ is the quotient of
$\SF$ by this ideal. The knowledge of the ideal $\FI \subset \SF$ (or
a set of generating elements) therefore determines $\Fh$. 

\vskip2mm{\bf 2.5. Additional structure \cite{PR2,P4}:}

A maximal subalgebra $\FA\subset\Fh$ of invariants with vanishing
Poisson brackets has been identified in \cite{PR2}. It is generated by
a Cartan subalgebra of the degree $0$ invariants and infinitely many
Lorentz (pseudo) scalars. The latter are obtained by varying the
generating functionals $\Tr (\log \phi_{\lambda\Gamma})^K$, where
$T^\mu=\lambda\Gamma^\mu$ and $\Gamma^\mu$ a matrix representation
of the Lorentzian Clifford algebra in $D$ dimensions, with respect to
the single parameter $\lambda$. Pseudoscalars arise only in odd
space-time dimensions.

The scalar generators $\BB_{(l)}$ belong to this abelian algebra $\FA$. 
The following remarkable feature has been observed. Let $\FU$ be the
subalgebra of $\Fh$ generated by the (finite) set of the remaining 
generators $J_1$, $S_1$, $S_2$, $T_2$. If $m^2>0$, the generators
$\BB_{(l)}$ can be modified to generators $\BB_0^{(l)}$ within $\FA$
in such a way that their linear span $\Fa\subset\FA$, acting by
Poisson brackets on $\FU$, takes $\FU$ into itself,   
\begin{equation}
\{\Fa,\FU\} \subset \FU.
\end{equation}
This observation has been verified in $d=1+3$ \cite{P3} and $d=1+2$
(unpublished) by explicit calculations up to considerable degree, but
a complete proof is lacking due to the highly non-trivial combinatorics 
involved.   

Assuming that these properties are confirmed at all degrees, then the 
Poisson ideal $\FI\subset\SF$, which determines the classical algebra
$\Fh$ as discussed in Sect.\ 2.4, acquires further structure. It is
generated by \cite{P4}  

$\bullet$ the Poisson ideal in the symmetric envelopping algebra of
the free Lie algebra with finitely many generators $J_1$, $S_1$, $S_2$,
$T_2$ which determines the Poisson algebra $\FU$. This ideal can be given
by a set of generating polynomial relations.

$\bullet$ the relations stating the Poisson commutativity of the
linear space $\Fa$.

$\bullet$ the commutative Poisson action of $\Fa$ on the Poisson algebra 
$\FU$, defining a semidirect product $\Fa \ltimes \FU \subset \Fh$.  
The corresponding ideal is generated by the Poisson brackets between
the basis elements $\BB_0^{(l)}$ of $\Fa$ and the generators $J_1$,
$S_1$, $S_2$, $T_2$ of $\FU$, taking values in $\FU$.
\vskip2mm

The details of the general structures outlined above may depend on the
space-time dimension $D$. They can be found in the given references.
The relevant ones will be introduced more explicitly in the next sections.

\section{Algebraic quantization: the general strategy}
\setcounter{equation}{0}

\indent\vskip2mm{\bf 3.1. The intrinsic approach to string quantization}

The intrinsic approach to the quantization of the string observables,
advocated by Pohlmeyer et al.\ \cite{P3,P4,HN}, consciously avoids all
reference to non-invariant monodromy variables $\RT$ and proceeds
directly from the Poisson algebra of classical {\em invariants} $\Fh$ 
or equivalently, from the Poisson ideal $\FI\subset\SF$ defining it. 

The polynomial nature of the relations generating the ideal $\FI$
gives rise to non-trivial self-consistency problems in the course
of quantization: the non-commutative quantum algebra should be the
quotient of the universal (i.e., non-commutative) envelopping algebra
$\UF$ of the free Lie algebra $\widehat\FF$ of the quantum generating
invariants by the ideal of the quantized polynomial relations. This
quantum ideal should exhibit a maximum of structural similarity with
the classical ideal. 

Thus, in order to define the quantum algebra, the infinite set of
generating quantized polynomial relations has to be specified in the
universal envelopping algebra $\UF$. These may differ from the classical
generating relations only by quantum corrections of lower degree,
expressible in terms of quantum generators (respecting spin and
parity). The quantum corrections are strongly constrained by the
postulate (principle of correspondence), that commutators between
quantum relations and quantum generators must not produce new quantum
relations (of order $\hbar$) without a classical counterpart. 

This, completely intrinsic, strategy has been pursued in refs.\
\cite{P3,P4,HN} and has been shown to be self-consistent in a highly
non-trivial manner up to degree five (in $1+3$ dimensions).

\vskip2mm{\bf 3.2. The new approach to string quantization}

We propose, in order to quantize $\Fh$, to quantize in a first step
the Lie algebra $\FR$ of monodromy variables by replacing the 
Poisson bracket (\ref{modpoiss}) by $i\hbar$ times the commutator, and
its symmetric envelopping algebra $\SR$ by the universal envelopping
algebra $\UR$ of the quantized Lie algebra $\HR$. The grading of $\SR$
by the degree does not survive the quantization because an evaluated
commutator has lower degree than the corresponding difference of
products (i.e., quantum corrections have lower degree). The remnant of
the grading $\SR=\bigoplus_l \SR^{(l)}$ is thus a {\em filtration}
\begin{equation}
\label{ufilter}
\UR=\bigcup_{l\geq-1} \UR^{(l)}, \qquad
\begin{array}{l}
\UR^{(l)}\cdot\UR^{(l')} \subset \UR^{(l+l'+1)} \\[2mm]
[\UR^{(l)},\UR^{(l')}] \subset \UR^{(l+l')}.
\end{array}
\end{equation}
$\UR^{(l)}$ is spanned by all non-commutative monomials of order $K$
in components of $\RT_\MMU$ with total tensor rank $N$ and $N-K-1\leq l$.  
The inclusions $\UR^{(l-1)} \subset \UR^{(l)}$ take care of the
ambiguities concerning the grading of quantum polynomials when
relations $[A,B]=AB-BA$ are taken into account.

The quantum algebra $\Hh$ of the invariants will be a subalgebra
of $\UR$ with a corresponding filtration through
$\Hh^{(l)}=\Hh \cap \UR^{(l)}$,
\begin{equation}
\label{filter}
\Hh = \bigcup_{l\geq-1} \Hh^{(l)}, \qquad
\Hh^{(l-1)} \subset \Hh^{(l)}, \qquad
\begin{array}{l}
\Hh^{(l)}\cdot\Hh^{(l')} \subset \Hh^{(l+l'+1)}
\\[2mm]
[\Hh^{(l)},\Hh^{(l')}] \subset \Hh^{(l+l')}
\end{array}.
\end{equation}

For the purpose of constructing this subalgebra $\Hh$, it is {\em not}
sufficient to specify the quantum counterparts of the {\em generating}
invariants as elements of $\UR$ such that they coincide with the classical
invariants in $\SR$ as the factor ordering and quantum corrections
of order $\hbar$ are ignored in $\UR$. The difficulty is due to the
polynomial relations among the quantum generating invariants (and
their commutators), which must hold in $\Hh$. Since $\Hh\subset \UR$,
they must be identities in $\UR$. But by construction, they hold in
$\UR$ only up to quantum corrections of order $\hbar$ or higher. The
principle of correspondence would be violated unless these quantum
corrections are also polynomials in the quantum generators. Otherwise
there would be elements of the quantum algebra without classical
counterpart. To implement this property required by the principle of
correspondence, using an ansatz for the quantum corrections to the
quantum generators, would require to actually compute the quantum
corrections to infinitely many polynomial relations. 

We circumvent this difficulty by defining the algebra of the
quantum invariants at one stroke as the subalgebra of $\UR$ which is
annihilated by a suitable derivation $\delta$:
\begin{equation}
\Hh := \Ker\delta \subset \UR
\end{equation}
where $\delta$ maps $\UR$ into another auxiliary algebra $\UC$. This
quantum condition is modelled after the classical condition
(\ref{repinv}), (\ref{classker}). The property of $\delta$ as a
derivation ensures that its kernel $\Ker \delta$ is an associative,
filtered {\em algebra} (Prop.\ 2). Then, it suffices to identify the
quantum counterparts of the classical generating invariants as
elements of $\Ker\delta$, and 
verify that they generate $\Ker\delta$ (the principle of correspondence,
Prop.\ 4). From this knowledge one concludes that the correction terms
in the quantum polynomial relations belong to $\Ker\delta$, and thus are
again quantum invariants, without the need to actually compute them.
The principle of correspondence will be established (Prop.\ 4) in the
form of linear isomorphisms between the classical spaces $\Fh^{(l)}$
and the quotients of the filtered spaces $\Hh^{(l)}/\Hh^{(l-1)}$.

These steps result in a consistent algebraic quantization of the 
repara\-metrization invariant observables of the closed bosonic string
in any space-time dimension. We shall show that they can be achieved
provided the quadratic generation property is true.

Explicit calculations \cite{M} in $1+3$-dimensional space-time show 
complete agreement with the intrinsic approach of ref.\ \cite{P4}
(with the intrinsically so far undetermined parameters taking specific
values). In particular, the additional structures outlined in Sect.\ 2.5,
$\{\FA,\FA\}=0$ and $\{\Fa,\FU\}\subset \FU$, also hold as commutator
relations in the quantum algebra $\Hh$ as far as they could be
checked, cf.\ also Sect.\ 7 and note added in proof.  

\section{Construction of the quantum algebra}
\setcounter{equation}{0}
We want to construct the quantum algebra of observables according to
the strategy just outlined. In order to focus our attention on the
relevant algebraic structures, we proceed from an abstract description of
the Lie algebra $\FR$ and the classical Poisson algebra $\Fh$ based on the
results of \cite{PR1}. It does not refer to the interpretation of the
algebra elements as functionals of the string's world surface, but
completely takes into account their algebraic aspects.
Along the way, we provide the necessary details omitted in Sect.\ 2.

Let $e_0,\ldots,e_{D-1}$ be a basis of $\RD$. A {\bf shuffle sum} is a
sum over basis vectors of $\otimes^N \RD$ of the form 
\begin{align}
\label{shuffle} {e^{\otimes N}}_
{\shuf{\mu_1\ldots\mu_K}{\mu_{K+1}\ldots\mu_N}} :=
\sum_{\substack{\pi \in \mathfrak{S}_N\\
    \pi^{-1}(1)<\cdots<\pi^{-1}(K)\\
    \pi^{-1}(K+1)<\cdots<\pi^{-1}(N)}} 
e_{\mu_{\pi(1)}}\otimes\ldots\otimes e_{\mu_{\pi(N)}} \quad \\[-8mm]
(0\leq K \leq N). \nonumber 
\end{align}
Let $V_N$ be the subspaces of $\otimes^N \RD$ spanned by all shuffle
sums with $1\leq K < N$, and $p_N:\otimes^N \RD\to \otimes^N \RD/V_N$
the canonical projections onto the quotient spaces $\otimes^N \RD/V_N$.
We identify the {\bf monodromy variables} with the components of the
tensors $\RT_\MMU=p_N(e_{\mu_1}\otimes\ldots\otimes e_{\mu_N})$,
i.e., they satisfy the defining linear relations
\begin{equation}
\label{linear}
\RT_{\shuf{\mu_1\ldots\mu_K}{\mu_{K+1}\ldots\mu_N}}=0 \qquad (1\leq K<N).
\end{equation}
Their linear span
$\FR=\bigoplus_{N=1}^\infty \otimes^N \BR^D/V_N$ equipped with
the bracket
\begin{align}
\label{modpoiss}
\{\RT_\mu,\RT_\nu\} &= \{\RT_\MMU,\RT_\nu\} = 0, \\
\{\RT_\MMU,\RT_{\nu_1\ldots\nu_M}\} &= 2\sum_{n=1}^N\sum_{m=1}^M
(-1)^{N-n+m}\eta_{\mu_n\nu_m}
\RT_{\shuf{\mu_1\ldots\mu_{n-1}}{\mu_N\ldots\mu_{n+1}}
\shuf{\nu_{m-1}\ldots\nu_1}{\nu_{m+1}\ldots\nu_M}}\;, \nonumber
\end{align}
is an infinite-dimensional Lie algebra, graded with respect to the degree
$l=N-2$. The second formula holds for $N\geq 2$, $M\geq 2$, and cannot
be extended to $N=1$ or $M=1$ without violating the Jacobi
identity. Instead, by the first formula, $\RT_\mu\equiv\PP_\mu$ are
central elements of $\FR$. 

The bracket canonically extends to the symmetric envelopping algebra
$\SR$ of $\FR$, which becomes a Poisson algebra. $\Fh\subset\SR$ is
the subspace spanned by the polynomials ({\bf invariants}) of the form
\begin{align}
\label{zyklpol}
\ZZ_\MMU &= \sum_{K=1}^N \ZZ^{(K)}_\MMU \\
\ZZ^{(K)}_\MMU &= \frac{1}{K!}\; \mathfrak{z}_N \circ
\bigg( \sum_{ 1\leq a_1<\ldots<a_{K-1}<N} \hbox{\kern-3mm}
\RT_{\mu_1\ldots\mu_{a_1}}
\RT_{\mu_{a_1+1}\ldots\mu_{a_2}}\cdots\RT_{\mu_{a_{K-1}+1}\ldots\mu_N}
\bigg)  \nonumber
\end{align}
where $\mathfrak{z}_N$ denotes the sum over the cyclic permutations
of the Lorentz indices. This space $\Fh$ is
in fact a Poisson subalgebra of $\SR$.

We capture the characterization of the invariants by their
reparametrization invariance, eqs.\ (\ref{ableit}) and (\ref{repinv}),
with the help of the linear map $\partial$ which maps $\FR$ into
$\RD\otimes\FR$ by
\begin{equation}
\label{partial}
\partial(\PP_\mu):=0,\qquad \partial(\RT_\MMU) :=
e_{\mu_1}\otimes\RT_{\mu_2\ldots\mu_N} - 
e_{\mu_N}\otimes\RT_{\mu_1\ldots\mu_{N-1}}.
\end{equation}
$\partial$ extends canonically to a derivation $\partial$ from the
commutative algebra $\SR$ into the commutative algebra $\SC$ (along
the injection homomorphism from $\SR$ into $\SC$). The invariants are
exactly those elements of $\FR$ which are annihilated by $\partial$:  
\begin{equation}
\label{kernel}
\Fh=\Ker(\partial)\subset \SR
\end{equation}
cf.\ (\ref{classker}).

This abstract description of the classical algebra of string observables
is the starting point for its quantization. As the monodromy variables
$\RT_\MMU$ form a Lie algebra, they can easily be quantized by defining
the commutator as $i\hbar$ times the Lie bracket $\{\cdot,\cdot\}$.
We call this Lie algebra $\HR$. By the theorem of
Poincar\'{e}-Birkhoff-Witt, the universal envelopping algebra $\UR$
is an associative filtered algebra. 

We want to parallel the prescription (\ref{kernel}) by ``quantizing''
the derivation $\partial$. But whereas a linear map defined on a Lie
algebra (considered as a linear space) extends canonically to a
derivation on its symmetric envelopping algebra, the same is not true
for an extension to the universal envelopping algebra, unless the map
has the property of a derivation with respect to the Lie bracket. 
For this reason, we first need an extension of the Lie bracket on $\HR$ to
a Lie bracket on the auxiliary space $\FC :=\RD \oplus\HR$ by defining
suitable commutation relations between the elements of $\HR$ and
$\RD$, with respect to which the ``quantized'' map $\delta: \HR\to
\UC$ is a derivation. Then $\delta$ extends to $\UR$, and we can
define the quantum algebra of observables $\Hh$ as its kernel within $\UR$.

\vskip2mm{\bf Proposition 1:}
1. The bracket $[\cdot,\cdot]: \FC \times \FC \to \FC$ on the
auxiliary space $\FC:=\RD\oplus\HR$
\begin{align}
\label{lie}
&[\RT_I,\RT_J]:=i\hbar\; \{\RT_I,\RT_J\}\;, & & [e_\mu,e_\nu] := 0 & \\
&[\RT_\MMU,e_\mu]:=0 \quad (N\neq 2), & & [\RT_{\mu\nu},e_\kappa]
:= -2i\hbar\; (\eta_{\nu\kappa}e_{\mu}-\eta_{\mu\kappa}e_{\nu}) & \nonumber
\end{align}
is a Lie bracket. As a Lie algebra, $\FC$ is a semidirect sum of
$\HR$ with the abelian Lie algebra $\RD$.

2. The linear map $\delta$ from $\HR$ into the universal envelopping
algebra $\UC$,
\begin{align}
\label{deltadef}
\delta(\RT_\mu) &:= 0 \\
\delta(\RT_\MMU) &:= \tfrac12[e_{\mu_1},\RT_{\mu_2\ldots\mu_N}]_+
- \tfrac12[e_{\mu_N},\RT_{\mu_1\ldots\mu_{N-1}}]_+\; (N\geq 2), \nonumber
\end{align}
where $[\cdot,\cdot]_+$ denotes the anti-commutator in
$\UC$, is well-defined and has the property of a derivation:
\begin{equation}
\delta([A,B])=[\delta(A),B]+[A,\delta(B)] \qquad
\forall \; A,B\in \FR\;.
\end{equation}
It extends canonically to a derivation $\delta: \UR \to \UC$ (along
the injection homomorphism).

\vskip2mm {\bf Comment:}
Whereas the map $\delta$ is obviously a quantum analogue of a
corresponding classical structure (the derivative $\partial_\sigma$,
eq.\ (\ref{ableit}), or equivalently the map $\partial$, eq.\
(\ref{partial})), the same is {\em not} true for the extension of 
the Lie bracket. The canonical bracket between $\RT_\MMU(\tau,\sigma)$
and $u_\mu(\tau,\sigma)$ is ill-defined and violates the Jacobi
identity. Therefore, the success of our quantum prescription is a
rather non-trivial feature.

\vskip2mm {\bf Proof:}
1. As the monodromy variables with the bracket $\{\cdot,\cdot\}$
form a Lie algebra, it is sufficient to prove the Jacobi identity for
multiple commutators involving the basis vectors of $\RD$. This can be
explicitly verified from the above definitions and the definition of
the Poisson bracket, eq. (\ref{modpoiss}).

2. $\delta$ is well-defined if it respects the linear dependencies
(\ref{linear}) among the monodromy variables. This is verified by
straightforward calculation using recursive relations for the shuffle
sums following from their definition (\ref{shuffle}):
\begin{align}
\label{rec}
0=\RT_{\shuf{\mu_1\ldots\mu_K}{\mu_{K+1}\ldots\mu_N}} &=
\RT_{\mu_1\shuf{\mu_2\ldots\mu_K}{\mu_{K+1}\ldots\mu_N}} +
\RT_{\mu_{K+1}\shuf{\mu_1\ldots\mu_K}{\mu_{K+2}\ldots\mu_N}}\nonumber \\
&=\RT_{\shuf{\mu_1\ldots\mu_{K-1}}{\mu_{K+1}\ldots\mu_N}\;\mu_K} +
\RT_{\shuf{\mu_1\ldots\mu_K}{\mu_{K+1}\ldots\mu_{N-1}}\;\mu_N}\;.
\end{align}
The derivation property of $\delta$ is obtained with a rather lengthy
but equally straightforward calculation \cite{M} from the definition
(\ref{lie}) of the Lie bracket together with (\ref{rec}). The case
$N=2$ has to be treated separately. 
\QED

As $\delta$ extends to a derivation on the universal envelopping
algebra $\UR$, products and commutators of any two elements in its
kernel are also annihilated by $\delta$. Hence its kernel $\Ker \delta$
is a subalgebra of $\UR$:

\vskip2mm {\bf Proposition 2:}
The kernel $\Hh=\Ker \delta \subset \UR$ of the derivation
$\delta$ is an associative non-commutative subalgebra of $\UR$.
\vskip2mm

It will be shown in the next section that $\Hh$ is a filtered algebra
and its filtration is compatible with the grading of the classical
algebra $\Fh$. Furthermore, it satisfies the principle of
correspondence in the sense of an association between classical
invariants and quantum invariants up to order $\hbar$, which respects
both products and brackets. 

\section{The Principle of Correspondence}
\setcounter{equation}{0}
As discussed in section 2, the classical Poisson algebras of monodromy
variables, $\SR = \bigoplus_l \SR^{(l)}$, and of invariants,
$\Fh = \bigoplus_l \Fh^{(l)}$, are graded with respect to the degree
$l\geq -1$, and
\begin{equation}
\Fh^{(l)}=\Ker \partial \cap \SR^{(l)}.
\end{equation}
Since the underlying Lie algebra for the commutative envelopping algebra 
$\SR$ is the same as for the non-commutative envelopping algebra $\UR$
(except for the factor $i\hbar$), the latter exhibits a corresponding
filtration $\UR=\bigcup_l \UR^{(l)}$ with $\UR^{(l-1)}\subset \UR^{(l)}$, 
cf.\ eq.\ (\ref{ufilter}). The subspaces $\UR^{(l)}$ are spanned by
all non-commutative monomials of order $K$ in components of $\RT_\MMU$
with total tensor rank $N$ and $N-K-1\leq l$. 

Due to the fact that a commutator $[A,B]$ is of lower degree than the
individual products $AB$, $BA$, the quotient spaces are naturally
isomorphic to the classical spaces of degree $l$ in $\SR$: 
\begin{equation}
\UR^{(l)}/\UR^{(l-1)}\cong \SR^{(l)}.
\end{equation}
The corresponding statements hold also for the commutative and
non-com\-mutative envelopping algebras $\SC$ and $\UC$, where we assign
the degree $l=-1$ to the generators of $\RD$. The derivation $\delta$
takes $\UR^{(l)}$ to $\UC^{(l-1)}$. 

We denote by $\Pi^{(l)}$ the canonical projection from $\UR^{(l)}$ to
$\SR^{(l)}$, and use the same symbol for its continuation to
$\UC^{(l)}$. In particular, applied to a non-commutative polynomial,
$\Pi^{(l)}$ is insensitive to the order of factors.
We have
\vskip2mm {\bf Proposition 3:} The following intertwining rules hold for all
$A\in\UR^{(l)}$, $B\in\UR^{(l')}$
\begin{align}
\label{multcorr}
\Pi^{(l+l'+1)}(A\cdot B) &= \Pi^{(l)}(A)\cdot \Pi^{(l')}(B), \\
\label{brackcorr}
\Pi^{(l+l')}([A,B]) &= i\hbar\; \{\Pi^{(l)}(A),\Pi^{(l')}(B)\}, \\
\label{dericorr}
\Pi^{(l-1)}\circ\delta(A) &= \partial \circ\Pi^{(l)}(A).
\end{align}

{\bf Proof:}
Due to the parallelism between the very definitions of the classical
and quantum operations (product, bracket, and derivation) on $\FR$
resp.\ $\HR$, their continuations to $\SR$ resp.\ $\UR$ differ only by
the necessity of factor ordering for the latter. Thus, ignoring factor
ordering and terms of lower degree (in particular commutators) by
means of the projections, takes the quantum operations to the
classical operations. 
\QED

Because of its definition as the kernel of a derivation, the quantum
algebra $\Hh$ of invariants inherits the filtration, as anticipated in
eq.\ (\ref{filter}):
\begin{equation}
\Hh=\bigcup_{l\geq-1} \Hh^{(l)},
\qquad\Hh^{(l)}=\Ker\delta \cap \UR^{(l)}\;.
\end{equation}

{\bf Corollary:}
The projections $\Pi^{(l)}$ map the spaces of quantum observables
$\Hh^{(l)}$ injectively into the spaces of classical observables
$\Fh^{(l)}$. Non-commutative multiplication and commutators are mapped
to commutative multiplication and Poisson brackets.

\vskip2mm {\bf Proof:}
By eq.\ (\ref{dericorr}), the image of $\Hh^{(l)} = \Ker \delta \cap \UR^{(l)}$
belongs to $\Ker \partial \cap \SR^{(l)}$ which equals $\Fh^{(l)}$. 
The projection is injective on $\Hh^{(l)}$ because it is injective on
$\UR^{(l)}$. The other statements are just reformulations of eqs.\
(\ref{multcorr}), (\ref{brackcorr}). 
\QED

Thus, if we show that $\Pi^{(l)}$ is also surjective onto $\Fh^{(l)}$,
we have linear isomorphisms between the classical and quantum
observables, preserving the pertinent algebraic structures up to
quantum corrections. Hence every element of $\Hh$ corresponds to a
classical invariant in leading order of $\hbar$.

\vskip2mm {\bf Proposition 4:}
1. For each classical invariant $\ZZ^{(2)}_\MMU$ of polynomial
order $K=2$, the corresponding quantum invariant defined by
\begin{equation}
\label{order2}
\widehat{\ZZ}^{(2)}_\MMU=\tfrac12 \;\mathfrak{z}_N\circ\bigg(\sum_{1\leq a<N}
\RT_{\mu_1\ldots\mu_a}\cdot\RT_{\mu_{a+1}\ldots\mu_N}\bigg)\;\in \UR^{(N-3)}
\end{equation}
lies in the kernel of derivation $\delta$.

2. Thus, by the quadratic generation hypothesis, all classical generators 
have their quantum counterparts in $\Hh$ given by eq.\ (\ref{order2}).

3. Using the generation of the classical algebra of observables, this
defines linear assignments $\alpha^{(l)}: \Fh^{(l)}\to\Hh^{(l)}$, such that
\begin{equation}
\label{surj}
\Pi^{(l)}\circ\alpha = {\rm id},
\end{equation}
and all elements of the form
\begin{align}
\alpha^{(l+l'+1)}(A\cdot B) - \alpha^{(l)}(A)\cdot\alpha^{(l')}(B)\\
\hbox{or}\qquad
\alpha^{(l+l')}(i\hbar\,\{A,B\})-[\alpha^{(l)}(A),\alpha^{(l')}(B)]
\end{align}
with $A\in \Fh^{(l)}$, $B\in \Fh^{(l')}$, are annihilated by
$\Pi^{(l+l'+1)}$ and $\Pi^{(l+l')}$, respectively. 

4. The projections $\Pi^{(l)}$ are surjective.

\vskip2mm {\bf Proof:}
Statement 1 is shown to be true by straightforward calculation using
the definition of the derivation $\delta$ and of the bracket
$[\cdot,\cdot]$ between $\RD$ and $\HR$. Statement 2 is obvious.

The assignment $\alpha$ is obtained as follows. Represent an element of
$\Fh^{(l)}$ as a polynomial in the generators and their multiple
Poisson brackets, i.e., as an element of the Poisson algebra $\SF$
(cf.\ Sect.\ 2.4), applying any convention regarding the addition of
polynomial relations (of the same degree) from the ideal $\FI$.
Replace in this polynomial every monomial by a corresponding product
of multiple commutators of the quantum generators according to the
previous statements, applying any convention regarding factor ordering. 
The resulting polynomial is an element of $\UR^{(l)}$, and because
$\delta$ is a derivation which annihilates the quantum generators,
this polynomial is also annihilated by $\delta$ and belongs to
$\Hh^{(l)}$. Applying the projection $\Pi^{(l)}$ restores the
classical original polynomial thanks to Prop.\ 3, proving eq.\
(\ref{surj}). In particular, all ambiguities due to the conventions
above are annihilated by the projection, because commutators in
$\Hh^{(l)}$ belong to $\Hh^{(l-1)}$, and because the polynomial
relations hold in $\Fh$. The other asserted properties of
$\alpha^{(l)}$ are then fulfilled by construction, because the two 
terms to be compared just correspond to two different conventions for
the assignment $\Fh^{(l)}\to\Hh^{(l)}$.

Eq.\ (\ref{surj}) implies surjectivity of $\Pi^{(l)}$ by
$\Fh^{(l)} =\Pi^{(l)}(\alpha^{(l)}(\Fh^{(l)}))$.
\QED

With Prop.\ 4, we have achieved the principle of correspondence for
the relation between the filtered quantum algebra $\Hh$ and the graded
classical algebra $\Fh$.

\vskip2mm {\bf Comment:}
The quantum invariants of order 2 are obtained as {\em symmetrized} quadratic
polynomials in the quantum monodromy variables. The same holds not
true for quantum invariants of higher order, since non-commutative
readings of eq.\ (\ref{zyklpol}) for $K > 2$ in general do not belong
to $\Ker\delta$. Instead, quantum invariants of higher order must be
defined through the generating property by the embeddings $\alpha^{(l)}$.

\section{The quadratic generation property}
\setcounter{equation}{0}

The assumed validity of the quadratic generation property of the classical
algebra $\Fh$ (cf.\ Sect. 2.4) enters our construction of the quantum
algebra only in Prop.\ 4. We want to emphasize that even if this 
hypothesis should fail, it would be sufficient for the construction if
for any other system of generating invariants the quantum counterparts
in $\Ker\delta$ could be given. 

By an argument given in \cite[Sect.\ IV]{PR1}, the quadratic
generation property would follow from a related property of the Lie
algebra $\FR$ (given below). There is considerable evidence for the
latter property to hold in any dimension, although partial general
arguments given so far are not yet conclusive \cite{P3,P5}. Explicit 
verification at least up to degree 7 is available in 1+3 dimensions
\cite[p.\ 28]{P4}.  

The relevant property of $\FR$ is most conveniently formulated in the
center of mass frame $\PP_\mu=(m,0,\dots,0)$ where there is an association
between an algebraic basis of $\Fh$ (``standard invariants'') and the
components of $\RT_\MMU$ with $\mu_1 \neq 0$, $\mu_N\neq 0$. The latter
elements of $\FR$ span a Lie subalgebra $\FR_{0}$. The ``exceptional''
elements $\sum_i\RT_{i0\dots0i}$ of odd rank $N=l+2$ cannot be generated 
by Poisson brackets within $\FR_{0}$ \cite[Sect.\ IV]{PR1}.%
\footnote{This circumstance explains the necessity of including, among
  the generators of $\Fh$, the scalars $\BB_{(l)}$ of odd degree $l$, 
  which have a non-vanishing ``leading'' contribution proportional
  to the exceptional elements.}
The hypothesis is that the elements of $\FR_{0}$ of degree $l\leq 1$
together with the infinite series of exceptional elements generate
$\FR_{0}$. Because the corresponding standard invariants are quadratic
\cite[Sect.\ IV]{PR1}, this property of $\FR_0$ entails the desired
quadratic generation property for $\Fh$.

Substantiating the expectation (spurred by extensive experience with
this algebra) that this property of the classical algebra holds in
general, would complete the construction of the quantum algebra in any
dimension. 

\section{Additional structures}
\setcounter{equation}{0}

We include some partial results concerning the persistence in the
quantum algebra $\Fh$ of the classical structures $\{\FA,\FA\}=0$ and
$\{\Fa,\FU\}\subset \FU$, mentioned in Sect.\ 2.5. Although the
survival of these additional structures in the quantum theory is highly
desirable, it is not a prerequisite for the quantization of the string
according to the prescription $\Hh = \Ker\delta$.

We shall prove the quantum commutativity of the quadratic elements of
the abelian subalgebra $\widehat\FA$ at any degree. These invariants are
obtained from the generating functional $\Tr(\log\Phi_{\lambda\Gamma})^2$
by variation with respect to the parameter $\lambda$, where the matrices
$\Gamma^\mu$ satisfy the Clifford algebra
$\Gamma^\mu\Gamma^\nu + \Gamma^\nu\Gamma^\mu = 2\eta^{\mu\nu}$. We
exploit the closed formula (\ref{modpoissgen}) for the Lie bracket on
$\FR$ which we expect to be a convenient tool for further structural study. 

The Clifford algebra implies the identity
\begin{equation}
\big[[\Gamma_\alpha,\Gamma_\beta]\otimes[\Gamma^\alpha,\Gamma^\beta],
\Gamma^\mu\otimes\Eins\big] =
8\big[\Gamma_\alpha\otimes\Gamma^\alpha, \Eins\otimes\Gamma^\mu\big].
\end{equation}
For $S^\mu=\lambda(\Gamma^\mu\otimes \Eins)$ and
$T^\mu=\kappa(\Eins\otimes\Gamma^\mu)$, this identity entails
\begin{equation}
[S\cdot T,S^\mu-T^\mu] = [W,S^\mu+T^\mu]
\end{equation}
with $W=\lambda\kappa\frac{\lambda^2+\kappa^2}{\lambda^2-\kappa^2}\,
(\Gamma_\alpha\otimes\Gamma^\alpha) -
\frac14\frac{\lambda^2\kappa^2}{\lambda^2-\kappa^2}\,
([\Gamma_\alpha,\Gamma_\beta]\otimes[\Gamma^\alpha,\Gamma^\beta])$.
By the latter identity, the (otherwise unwieldy) differential term in
the expression (\ref{modpoissgen}) for the bracket between the
generating functionals becomes
\begin{align}
\Tr\big([W,X]\cdot\partial_{X}\big)\log\Phi_X\vert_{X=S+T} & =
[W,\log \Phi_X\vert_{X=S+T}] \nonumber \\ & = [W,\log\Phi_S+\log\Phi_T], 
\end{align}
and (\ref{modpoissgen}) acquires the form without a differential term
\begin{align}
\label{modpoisscliff}
\{\log\Phi_{\lambda\Gamma}&\substack{\otimes\\,}\log\Phi_{\kappa\Gamma}\}
=\\ &=
[W,\log \Phi_{\lambda\Gamma}\otimes\Eins +
\Eins\otimes\log\Phi_{\kappa\Gamma}] -
[V,\log \Phi_{\lambda\Gamma}\otimes\Eins -
\Eins\otimes\log\Phi_{\kappa\Gamma}] \nonumber \end{align}
with $V=\lambda\kappa\,(\Gamma_\alpha\otimes\Gamma^\alpha)$, and $W$ as above.

The expansion of the generating functional $\log\Phi_{\lambda\Gamma}$
simplifies considerably, because by virtue of the shuffle symmetries
of the tensors $\RT_\MMU$, only multiple commutators of Clifford matrices
contribute to $\log\Phi_{\lambda\Gamma}$, which can be worked out
\cite[Sect.\ II]{PR3}:
\begin{equation}
\log\Phi_{\lambda\Gamma} = \BP_\mu(\lambda)\,\Gamma^\mu +
\BR_{\mu\nu}(\lambda)\,\tfrac12[\Gamma^\mu,\Gamma^\nu],
\end{equation}
where $\BP_{\mu}$ and $\BR_{\mu\nu}$ are certain $\FR$-valued power
series in the parameter $\lambda$, involving only odd and even powers
of $\lambda$, respectively. Hence 
\begin{equation}
\label{logphi2}
\Tr (\log\Phi_{\lambda\Gamma})^2 =
\BP_\mu(\lambda)\BP^\mu(\lambda)
-2 \BR_{\mu\nu}(\lambda)\BR^{\mu\nu}(\lambda).
\end{equation}

The Poisson bracket (\ref{modpoisscliff}) between
$\log\Phi_{\lambda\Gamma}$ and $\log\Phi_{\kappa\Gamma}$  
can now be used to derive the brackets among the power series
$\BP_\mu$ and $\BR_{\mu\nu}$. They read (after quantization)
\begin{align}
\label{prcomm}
[\BP_\mu(\lambda),\BP_\nu(\kappa)] & = -8i\hbar\,
\lambda^3\kappa^3\;
\frac{\lambda^{-2}\BR_{\mu\nu}(\lambda)-\kappa^{-2}\BR_{\mu\nu}(\kappa)}
{\lambda^2-\kappa^2}\\
[\BP_\tau(\lambda),\BR_{\mu\nu}(\kappa)] & = -2i\hbar\, \eta_{\tau\mu}\,
\lambda^3\kappa^2\;
\frac{\lambda^{-1}\BP_\nu(\lambda) -\kappa^{-1}\BP_\nu(\kappa)}
{\lambda^2-\kappa^2} - (\mu\leftrightarrow\nu) \nonumber \\
[\BR_{\sigma\tau}(\lambda),\BR_{\mu\nu}(\kappa)] & = -2i\hbar\,\eta_{\tau\mu}\,
\lambda^2\kappa^2\;
\frac{\BR_{\sigma\nu}(\lambda)-\BR_{\sigma\nu}(\kappa)}{\lambda^2-\kappa^2} -
(\mu\leftrightarrow\nu) - (\sigma\leftrightarrow\tau). \nonumber \end{align}
(The denominators always divide the power series in the numerator.)

Now, the quantum commutator between $\Tr(\log\Phi_{\lambda\Gamma})^2$
and $\Tr(\log\Phi_{\kappa\Gamma})^2$ given by eq.\ (\ref{logphi2}),
can be evaluated with eq.\ (\ref{prcomm}). One finds that the
resulting cubic terms again arrange into commutators, which can in
turn be evaluated with eq.\ (\ref{prcomm}). The resulting quadratic
terms once more arrange into commutators, which can be evaluated and
finally vanish.

We conclude that the infinite series of quantized invariants generated by
$\Tr (\log\Phi_{\lambda\Gamma})^2$ commute among each other.

Unfortunately, this kind of argument fails for higher powers of
$\log\Phi_{\lambda\Gamma}$, because the traces
$\Tr (\log\Phi_{\lambda\Gamma})^K$ do not belong to the kernel of
$\delta$, and the general form of the necessary quantum corrections is
not clear to us. The reason for the complication comes from the
commutator in
\begin{equation}
\delta \log\Phi_{\lambda\Gamma} = \lambda
[e_\mu\Gamma^\mu,\log\Phi_{\lambda\Gamma}]
\end{equation}
which is both with respect to the Clifford matrices and with respect to
the quantum operators in $\FC$, so that the matrix trace of a
commutator does not vanish. Yet, we expect that there should be a
likewise simple argument as in the classical case, giving rise to an 
equally large infinite system of commuting quantum observables.

The case of the semidirect action of $\Fa$ on $\FU$ (cf.\ Sect.\ 2.5)
appears much more difficult since it is still not well understood even
at the classical level. It has been verified in the quantum algebra
by explicit calculation in $1+3$ dimensions up to degree 4 \cite{M} but a
general proof is still lacking both for the quantum and the classical
case (see note added in proof).

\section{Conclusion}
\setcounter{equation}{0}

We have established a rather simple characterization of the quantum
algebra of the gauge invariant observables of the closed bosonic
string in any dimension, as the kernel of a suitable derivation
\begin{equation}
\Hh = \Ker \delta \subset \UR.
\end{equation}
$\delta$ is defined on the universal envelopping algebra $\UR$ of an
infinite-dimens\-ional Lie algebra $\HR$ with values in $\UC$ where
$\FC$ is a Lie algebra extension of $\HR$. The only possible loophole
in the argument is the lacking proof of an apparent structural
property of the underlying {\em classical} theory, which is required
for the proper correspondence with the quantum algebra.

In Sect.\ 7, we gave some first results which indicate that also
the ``additional structures'' mentioned in Sect.\ 2.5 are respected by
quantization.

The present algebraic definition of the quantum algebra bears some obvious
resemblance with the BRST quantization scheme, where the derivation
$\delta_{\rm BRST}$, given by the commutator with a BRST charge
$Q$, determines the gauge invariant observables as its kernel.
In the case at hand, however, $\delta$ is given just as a
derivation which is not implemented as a commutator. This suffices
to ensure that $\Hh$ is an algebra. (It is also a * algebra, if one
treats the generators of the Lie algebras $\HR$ and $\FC$ as hermitean.)

We consider the quantized ambient algebras $\UR$ and $\UC$ only as 
auxiliary tools in order to establish the existence of $\Hh$ with the
desired properties. A representation of $\Hh$ does not presuppose a
representation of $\HR$, and in particular no ``charge'' operator
which would implement $\delta$. We doubt the existence of such an
object since $\delta$ as a map from $\UR$ to $\UC$ has no sensible
extension to its image. It is likewise not meaningful to talk about
$\delta$-invariant states. 

The search for Hilbert space representations of the quantum string is
therefore an independent problem. It should not be biased by any
features of the ambient algebra $\UR$ unless they are intrinsic 
to $\Hh$. The construction (classification) of representations of
$\Hh$ is an exciting challenge for the future.  

\vskip2mm{\bf Acknowledgements.}
One of the authors (C.M.) wishes to thank Prof.\ K. Pohlmeyer, supervisor of
her diploma thesis at the Physics Department of Freiburg University. Both of
us are grateful for his critical comments on a preliminary version of
the manuscript. 

\vskip2mm{\bf Note added in proof.} 
A failure of the semidirect product structure
$[\mathfrak{a},\mathfrak{U}]\subset \mathfrak{U}$ in the intrinsic
approach to quantization was reported recently in \cite{HPTW}.

\end{document}